\documentclass[twoside]{article}
\usepackage{graphicx}
\usepackage[skip=10pt]{parskip}

\usepackage{etoolbox}
\apptocmd{\thebibliography}{\interlinepenalty=10000}{}{}

\usepackage{caption}
\usepackage{cite} %
\usepackage{amsmath}
\usepackage{amsfonts}
\usepackage{bm}
\usepackage{multirow}
\usepackage{graphicx}
\usepackage{booktabs}
\usepackage{xcolor}
\usepackage{xspace}
\usepackage{hyperref}
\usepackage[capitalise]{cleveref}
\usepackage{authblk}
\usepackage{fancyhdr}
\newcommand{\custombullet}{\raisebox{0.3ex}{\scalebox{0.7}{\(\bullet\)}}}

\begin{document}
\title{\Large \textbf{RapidVol: Rapid Reconstruction of 3D Ultrasound Volumes from Sensorless 2D Scans}\\[0.5em]}
\fancyhead[RO]{RapidVol}

\fancyhead[LE]{M. Eid et al.}

\author[1,2,*]{Mark C. Eid}
\author[2]{Pak-Hei Yeung}
\author[2]{Madeleine K. Wyburd}
\author[1]{João F. Henriques}
\author[2]{Ana I.L. Namburete}
\renewcommand\Authfont{\large}

\affil[1]{Visual Geometry Group, University of Oxford}
\affil[2]{Oxford Machine Learning in NeuroImaging Lab, University of Oxford}
\affil[*]{\href{mailto:markeid@robots.ox.ac.uk}{markeid@robots.ox.ac.uk}}
\renewcommand\Affilfont{\normalsize}

\date{} %
\maketitle %
\begin{abstract}
\noindent
Two-dimensional (2D) freehand ultrasonography is one of the most commonly used medical imaging modalities, particularly in obstetrics and gynaecology. However, it only captures 2D cross-sectional views of inherently 3D anatomies, losing valuable contextual information. As an alternative to requiring costly and complex 3D ultrasound scanners, 3D volumes can be constructed from 2D scans using machine learning. However this usually requires long computational time. Here, we propose RapidVol: a neural representation framework to speed up slice-to-volume ultrasound reconstruction. We use \textit{tensor-rank decomposition}, to decompose the typical 3D volume into sets of tri-planes, and store those instead, as well as a small neural network. A set of 2D ultrasound scans, with their ground truth (or estimated) 3D position and orientation (pose) is all that is required to form a complete 3D reconstruction. Reconstructions are formed from real fetal brain scans, and then evaluated by requesting novel cross-sectional views. When compared to prior approaches based on fully implicit representation (e.g. neural radiance fields), our method is over 3x quicker, 46\% more accurate, and if given inaccurate poses is more robust. Further speed-up is also possible by reconstructing from a structural prior rather than from scratch.\\

\end{abstract}
\thispagestyle{empty} %

\newpage
\section{Introduction} \label{sec:Introduction}
Two-dimensional (2D) freehand ultrasonography is routinely used in prenatal checkups, as well as to image other organs. It is affordable, provides instant visualisation, can be repeated numerously due to lack of ionising radiation, and can even be carried out from a smartphone. However, whilst magnetic resonance imaging (MRI) and computerised tomography (CT) scans capture and store inherently 3D structures within the body as 3D volumes, the same is not true for freehand ultrasound (US). Three-dimensional (3D) US scans do exist, and have several clinical benefits over 2D methods, such as improved detection of cleft lip \cite{campbell_prenatal_2007,goncalves_three-dimensional_2016,chen_prenatal_2001}, and providing greater diagnostic accuracy irrespective of sonographer experience \cite{huang_review_2017,pistorius_grade_2010}. One study quantified this as a 60.8\% improvement \cite{merz_advantages_2017}. To acquire these 3D US scans, native 3D scanners can be used, however they are still not routinely used as they are $\sim$10 times more expensive than standard US probes and are bulkier \cite{yeung_implicitvol_2021}. Operators also require additional training.

An alternative to moving to 3D scanners, is to reconstruct 3D volumes from 2D freehand US scans, acquired from standard sensorless probes. Yeung. et al. \cite{yeung_implicitvol_2021} proposed a framework, ImplicityVol, to implicitly construct a 3D volume of a fetal brain using a series of freehand 2D US scans, which can also be applied to scans of other fetal or adult organs. Currently, ImplicitVol \cite{yeung_implicitvol_2021} takes $O(\text{hours})$ to reconstruct a 3D brain, however ideally this would be of $O(\text{minutes})$ so that acquisition \emph{and} reconstruction (as well as analysis of it by a clinician) can take place together within the same appointment.
\textbf{This paper therefore presents a new representation method, termed RapidVol, which is significantly more accurate (up to 46\%), and forms the reconstructed 3D brain from inputted 2D freehand ultrasound scans 3 times quicker}.

\section{Preliminary} \label{sec:Related Work}

Traditionally, 3D scenes are represented purely \textit{explicitly}, where the entire 3D volume/scene is stored as a grid of voxels \cite{seitz_photorealistic_1997,szeliski_stereo_1998}. More lightweight versions can use sparse or octree voxel grids, or meshes \cite{shalma_review_2023}. A modern technique is to instead store the scene \textit{implicitly}, by saving a fully trained neural network \cite{mildenhall_nerf_2020,flynn_deepview_2019,henzler_single-image_2018,zhou_stereo_2018}. More recently, hybrid methods have been proposed, combining and drawing upon the benefits of both pure explicit and implicit representation \cite{chen_tensorf_2022,chan_efficient_2022,tang_compressible-composable_2022,yu_plenoxels_2021}.

\textbf{Explicit Representation: }The 3D brain can be represented discretely as a cost volume $\boldsymbol{V} \in R^{H \times W \times D \times C}$, a tensor of height $H$, width $W$, depth $D$ and channels $C$. A 3D US volume will have $C=1$ as US is a grayscale modality, colour objects will have $C=3$. If there are any voxels which were not seen in any of the 2D images used for reconstruction, tri-linear interpolation, nearest neighbour, or spline fitting can be used to predict these missing voxel values \cite{antonio_aceves-fernandez_survey_2019,chen_mvsnerf_2021}.

\textbf{Implicit Representation:} Based of NeRF \cite{mildenhall_nerf_2020}, ImplicitVol \cite{yeung_implicitvol_2021} instead represents the 3D brain volume $\boldsymbol{V} \in \mathbb{R}^{H \times W \times D \times 1} $ as a trainable deep neural network, $F_\Theta : \mathbf{x} \rightarrow c$, where $\Theta$ denotes the learnable parameters, $\mathbf{x}$ the $(x,y,z)$ co-ordinates of any point within the 3D brain, and $c$ the intensity value.
By predicting intensity $c$ at a given input voxel location $\mathbf{x}$, the discrepancy between $c$ and the actual intensity at $\mathbf{x}$ can be computed as the loss to then update $\Theta$ through back-propagation.

    \begin{figure}[b]
    \centering
    \includegraphics[width=1\textwidth]{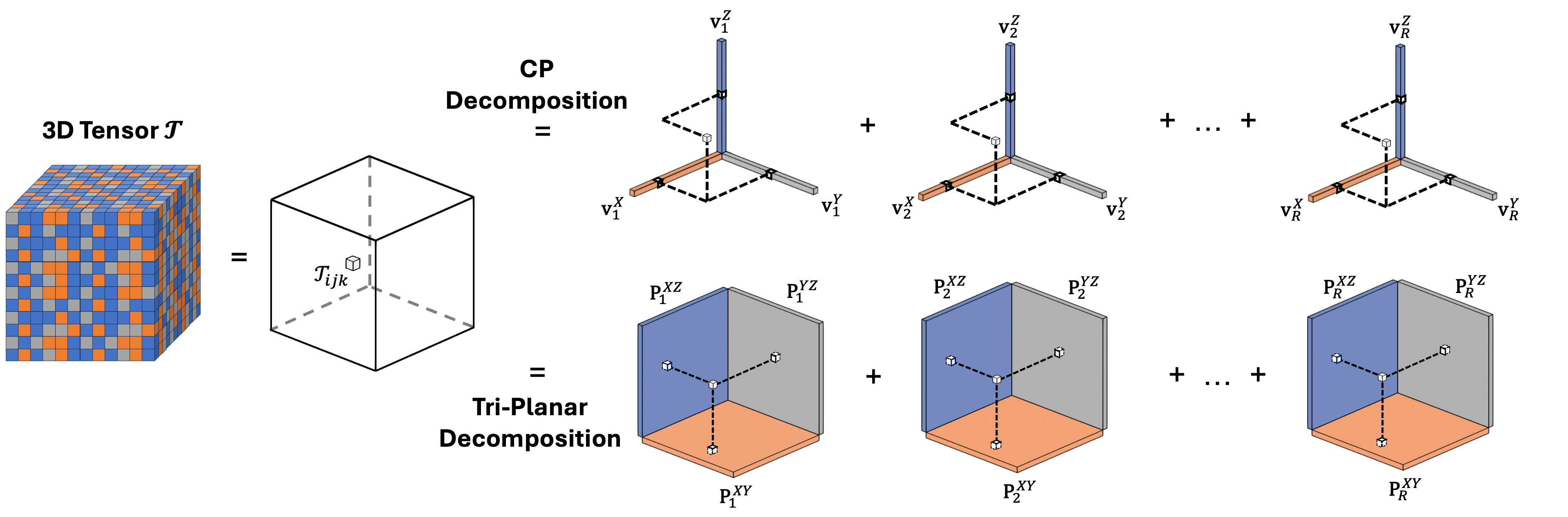}
    \caption{Diagram showing the two different types of tensor decomposition employed, and how the value of a voxel in $\mathcal{T}$ can be retrieved.}
    \label{fig:Decomposition}
    \end{figure}

\textbf{Hybrid Implicit-Explicit Representation:}
Implicit and explicit representation models have contrasting benefits and limitations. For example, explicit models scale \textit{cubically} with resolution, but are quicker to train. Recent works such as TensoRF \cite{chen_tensorf_2022} and EG3D \cite{chan_efficient_2022} attempt to bridge these two approaches and harness the benefits of both. As shown in \cref{fig:Decomposition}, they use tensor rank decomposition to instead store the $H \times W \times D \times C$ cost volume much more compactly as $C$ sets of three 1D Vectors (CP Decomposition \cite{carroll_analysis_1970}), or $C$ sets of three 2D planes (``Tri-Planar Decomposition") as in EG3D. Only a very small MLP is then needed to decode from $C$ channels to 1/3 colour channel(s). By using a mix of explicit and implicit representation techniques, storage reduces from $O(n^3)$ to $O(n^2)$ and $O(n)$ for Tri-Planar and CP Decomposition respectively. More importantly, the speed benefits associated with explicit representation are still maintained, and in CP Decomposition it is improved (as for each training image, $(H + W)$  rather than $(H \times W)$ parameters have to now be updated). Linear and bilinear interpolation can also be quickly applied within each individual 1D vector (CP Decomposition) or 2D plane (Tri-Planar Decomposition). This compensates for the lack of continuity when using pure explicit representation.

We thus adapt TensoRF \cite{chen_tensorf_2022} and EG3D \cite{chan_efficient_2022} from the original natural scene setting to the medical imaging setting, specifically on fetal brains. Challenges that must be addressed include the relative lack of training data, the much higher intricacies present in the brain as opposed to the Synthetic-NeRF dataset \cite{mildenhall_nerf_2020}, and the need for high accuracy and interpretability for clinical deployment.

\section{Methodology} 
\label{sec:Methodology}

\subsection{Problem Setup} \label{sec:Problem Setup}

\begin{figure}[h]
    \centering
    \includegraphics[width=1\textwidth]{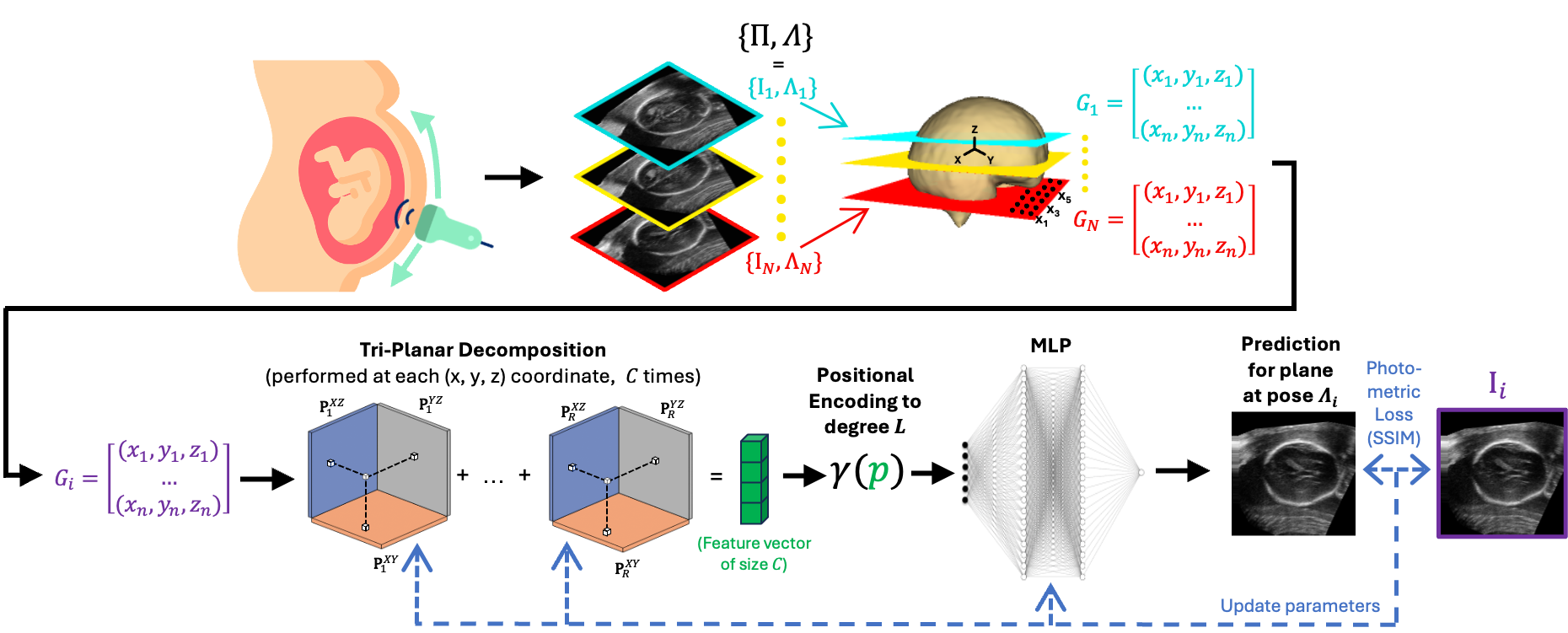}
    \caption{Pipeline of our proposed method RapidVol. During the reconstruction process, a set of $\Pi$ images and their corresponding poses $\Lambda$ are required as input. Once the brain is reconstructed, only the pose $\Lambda$ at which one wishes to see is required as input, and parameters are not updated.\\ \footnotesize{\textit{Nb. Ultrasound probe image adapted from Flaticon.com}.}}
    \label{fig:RapidVol Pipeline}
    \end{figure}

In a medical setting, we have a stack $\Pi$ of $N$ 2D ultrasound images \mbox{($\Pi=\{\mathbf{I}_i\}^N_{i=1}$)}, all of which are different cross-sections but of the same inherently 3D fetal brain (see \cref{fig:RapidVol Pipeline}). Each cross-sectional image has a known pose $\Lambda_i$ relative to the centre of the 3D brain, which can be parameterised by 3 Euler angles ($\mathbf{E}$) and 3 Translations ($\mathbf{T}$). Our goal is to reconstruct the 3D brain such that a 2D cross-section can be viewed at any specified pose $\Lambda$, and at any resolution. The final output of RapidVol \textit{appears} to be a high-resolution cost volume $\boldsymbol{V}$, however it is actually a set of tri-planes or tri-vectors, and a small MLP.

\subsection{Tensor Decompositions} \label{sec:Tensor Decomposition}
\textbf{Tri-Planar Decomposition:}
Given a 3D Tensor $\bm{\mathcal{T}}$ $\in$ $\mathbb{R}^{I \times J \times K} $, Tri-Planar Decomposition factorises $\mathcal{T}$ such that for a given $ijk$ index position within $\mathcal{T}$:
\begin{equation}
    \mathcal{T}_{ijk} = \sum^R_{r=1} \mathbf{P}^{XY}_{r,ij} \circ \mathbf{P}^{YZ}_{r,jk} \circ \mathbf{P}^{XZ}_{r,ik}
    \label{eq:Tri-Planar Decomposition index}
\end{equation}
where $\circ$ is either the sum (shown in \cref{fig:Decomposition}) or product. In preliminary results, we found that the product performs better, so exclusively used it throughout this work.
$\mathbf{P}^{XY}_{r} \in \mathbb{R}^{I \times J}$, $\mathbf{P}^{YZ}_{r} \in \mathbb{R}^{J \times K}$, $\mathbf{P}^{XZ}_{r} \in \mathbb{R}^{I \times K}$ are the factorised 2D planes for the $r^\text{th}$ component. $R$ is the rank of the decomposition, and is a user-selected hyper-parameter. Increasing $R$ improves the accuracy of the decomposition, but at the expense of memory and computation time.
Explicit representation utilises a 4D Tensor $\boldsymbol{V} \in R^{H \times W \times D \times C}$ with $C \geq 1$. However since $C<<{H,W,D}$, for simplicity the Channel dimension is not decomposed and instead \cref{eq:Tri-Planar Decomposition index} is repeated $C$ times. \cref{eq:Tri-Planar Decomposition index} can also be evaluated at non-integer $ijk$ indices by bilinearly interpolating within each tri-plane.

\subsubsection{CP Decomposition:} Given the same 3D Tensor $\bm{\mathcal{T}}$ $\in$ $\mathbb{R}^{I \times J \times K} $, CP Decomposition factorises $\mathcal{T}$ such that for a given $ijk$ index:
\begin{equation}
    \mathcal{T}_{ijk} = \sum^R_{r=1} \mathbf{v}^X_{r,i} \ \mathbf{v}^Y_{r,j} \ \mathbf{v}^Z_{r,k}
    \label{eq:CP Decomposition index}
\end{equation}
where $\mathbf{v}_r^X \in \mathbb{R}^I ,\mathbf{v}_r^Y \in \mathbb{R}^J,\mathbf{v}_r^Z \in \mathbb{R}^K$ are the factorised 1D vectors for the $r^\text{th}$ component. As before, $R$ is the rank and is user-selected, and \cref{eq:CP Decomposition index} is repeated for each Channel in $\boldsymbol{V}$. Non-integer indices can also be used by linearly interpolating. Both \cref{eq:Tri-Planar Decomposition index,eq:CP Decomposition index} can be visualised in \cref{fig:Decomposition}.

\subsection{Pipeline of RapidVol} \label{sec:Pipeline of RapidVol}
To form $\boldsymbol{V}$ from a set of 2D ultrasound freehand scans ($\Pi$) with known poses $\Lambda$, RapidVol undergoes the following steps (as shown in \cref{fig:RapidVol Pipeline}):

{\parskip=0pt\parindent=15pt %
\textbf{1.} The pose (parameterised by $\mathbf{E}$ and $\mathbf{T}$) of each image is used to construct a grid $G = \{\mathbf{x}_i\}_{i=1}^n = \{x_i,y_i,z_i\}_{i=1}^n$ of a nominated resolution, which contains the 3D co-ordinates of all $n$ pixels which lie on that cross-sectional plane/image.
    
\textbf{2.} $G$ is fed into our reconstruction model, which performs either \cref{eq:Tri-Planar Decomposition index} or \cref{eq:CP Decomposition index} on each $\mathbf{x}_i$. This in turn produces a grid of the same shape as $G$, but now with $C$ channels rather than 1. Positional encoding (see \hyperref[sec:Appendix A]{Appendix A} or \cite{mildenhall_nerf_2020}) is then applied on this grid, before being fed into the MLP.
    
\textbf{3.} A lightweight, trainable MLP decodes the positionally encoded grid of $C$ channels to 1 channel, resulting in a grayscale image of the fetal brain at the requested pose.
    
\textbf{4. } The loss, chosen to be the negative Structural Similarity Index Measure (SSIM) \cite{wang_image_2004} between the rendered image and the ground truth image, is computed. Using standard back-propagation the tri-planes (or tri-vectors) are then refined, as is the decoder MLP and optionally the poses of the training images.}

Once the brain is reconstructed, cross-sectional views at any pose can be viewed simply by specifying the pose and performing steps 1-3 above.

\section{Experimental Setup} \label{sec:Experimental Setup}
\subsection{Technical Details}
\textbf{Dataset:} Throughout, we use a set of 3D ultrasound fetal brain scans at 20 gestational weeks of size $160^3$ voxels, with a resolution of 0.6 mm$^3$ isotropic, collected as part of the INTERGROWTH-21st study \cite{papageorghiou_international_2014}. In practice, only a set of $N$ freehand 2D images are required for reconstruction. However to evaluate the accuracy of our reconstruction method, we also require a 3D scan of the fetus so that we can compare an arbitrary plane from the reconstructed volume to the ground truth (extracted from the 3D scan). For each fetus we only had a 3D scan, rather than a 3D scan \textit{and} a series of freehand images, so we had to mimic the latter. This was done by sampling $N$ linearly spaced axial images from the 3D scan, which simulates the ultrasound probe moving along the stomach from head to toe during a typical prenatal scanning session. $\Pi$ was then those $N$ images, and $\Lambda$ was their (ground truth) poses. When evaluating the accuracy of our reconstruction, views at those poses were naturally not requested.
 
\textbf{Implementation Details:} We implement our framework in PyTorch. Unless otherwise stated, the tri-planes/tri-vectors are randomly initialised. A Stochastic Gradient Descent (SGD) optimiser with a learning rate (lr) of 0.5 is used to refine the decomposition/recon\-struction model, and a SGD with a lr of 0.001 for the decoder MLP. Training is done for 5,000 epochs.

\subsection{Experiments Performed}\label{sec:Experiments Performed}
\textbf{Ablation Study:} We first performed an ablation study to find the optimum type of decomposition (Tri-Planar or CP), MLP (see \cref{fig:RapidVol Pipeline}), as well as the degree to positionally encode up to, and whether the raw input should be concatenated with the encoded input or not. The MLPs ablated over all had input size $C$, output size 1, and were two to four layers, with hidden layers being of size \{32,64,128\}. ReLU activation functions were used, except on the final layer which had a sigmoid activation function to constrain the output to be a grasyscale value between 0.0 and 1.0. 

\textbf{Performance Comparison:} We then compared the performance, both in terms of reconstruction accuracy and speed, of our method to a state-of-the-art reconstruction method, ImplicitVol \cite{yeung_implicitvol_2021}, on 15 3D fetal brains. Two different sets of input scans $\Pi$ were used: The first was a set of $N=\{128,256\}$ axial slices as previously described ($\Pi_1$), the second was a set of $N=\{128,256\}$ coronal slices uniformly rotated $360^{\circ}$ about the Vertical Axis of the brain ($\Pi_2$). This is to simulate the ultrasound probe being rotated by hand as is often done in a prenatal scan, and is the same dataset used in \cite{yeung_implicitvol_2021}. Reconstruction accuracy was quantified by requesting views of $N=\{128,256\}$ linearly spaced axial, coronal and sagittal slices. 

\textbf{Training from an Atlas:} Additionally, we investigated whether the reconstruction of a 20 week old fetal brain could be sped up by initialising the tri-planes/tri-vectors from an atlas of the same age, rather than from random values. This atlas was a pre-computed reconstruction of multiple fetal brains. All fetuses were of the same age, and the 3D scans of their brains were deemed to be of high enough quality to be used in constructing a digital atlas for that gestational week \cite{namburete_normative_2023}.

\textbf{Use of Inaccurate Poses:} Ordinarily, we mimicked the set of input freehand scans $\Pi$ by sampling $N$ images from a 3D scan, and so we knew their ground truth poses which is what we set $\Lambda$ to be. However in practice, the poses of the freehand images are unknown, so we require methods like PlaneInVol \cite{yeung_learning_2021} to predict them. Unfortunately neither the dataset used to train PlaneInVol nor a fully trained version of PlaneInVol is publicly available. In this experiment, we instead set $\Lambda$ to be the ground truth poses plus some random noise ($X{\sim}U(-3,3)$), to simulate the inevitably inaccurate pose estimation of PlaneInVol. The poses were also set to be learnable, so as training progressed, the poses could tend towards their ground truth values and reconstruction became more accurate. The poses were refined using an Adam optimiser with a learning rate of 0.001. Optimisation of the poses was done jointly with optimisation of the tri-planes and MLP, as in \cite{yeung_implicitvol_2021}.

\section{Results and Discussion}
\textbf{Ablation Study:}
CP decomposition, although quicker and more memory efficient than Tri-Planar Decomposition, was found to be \textit{too} simple to accurately reconstruct the intricate 3D fetal brain (see \hyperref[sec:Appendix B]{Appendix B}). A two-layer MLP with a hidden layer size of 64 (as used in \cite{muller_instant_2022}), alongside positional encoding to degree $L=2$ and concatenating the raw input with the encoded input, performed the best (see \hyperref[sec:Appendix B]{Appendix B}). For this MLP, it was found that the best compromise between accuracy, speed and memory storage was when Tri-Planar Decomposition was performed with $R=5$ and $C=10$. Therefore all subsequent experiments used Tri-Planar Decomposition with these parameters.

\begin{table}[b]
\caption{Evaluation of -SSIM accuracy results at test time for our method (RapidVol) vs the baseline (ImplicitVol) on 15, 20 week fetal brains. More negative SSIM is better.}\label{tab:Accuracy comparison}
\centering\scriptsize\setlength{\tabcolsep}{1pt}
\begin{tabular}{cccccccc}
\toprule 
 &  & \multicolumn{3}{c}{N = 128 Testing Slices} & \multicolumn{3}{c}{N = 256 Testing Slices}\\
\cmidrule{3-5} \cmidrule(l){6-8}
$\Pi$ & Method & Axial & Coronal & Sagittal & Axial & Coronal & Sagittal\\
\midrule 
\multirow{2}{*}{Axial} & RapidVol & N/A & -0.955\textpm 0.008 & -0.952\textpm 0.009 & N/A & -0.973\textpm 0.005 & -0.969\textpm 0.006\\
\cmidrule{2-8} 
 & ImplicitVol & N/A & -0.689\textpm 0.038 & -0.700\textpm 0.046 & N/A & -0.677\textpm 0.040 & -0.662\textpm 0.037\\
\midrule 
\multirow{2}{*}{\shortstack{\\$360^\circ$\\Coronal}} & RapidVol & -0.941\textpm 0.016 & -0.932\textpm 0.018 & -0.935\textpm 0.017 & -0.952\textpm 0.015 & -0.950\textpm 0.016 & -0.950\textpm 0.014\\
\cmidrule{2-8} 
 & ImplicitVol & -0.909\textpm 0.021 & -0.901\textpm 0.023 & -0.905\textpm 0.022 & -0.912\textpm 0.020 & -0.904\textpm 0.023 & -0.908\textpm 0.022\\
\bottomrule
\end{tabular}
\end{table}

\par{\noindent \textbf{Performance Comparison:}}
\cref{tab:Accuracy comparison} shows the reconstruction performance of RapidVol compared to current state of the art, ImplicitVol. If $\Pi$ is a series of scans acquired by rotating an US probe by hand ($\Pi_2$), then our method is more accurate than ImplicitVol at generating novel cross-sectional views, but only marginally so. Nevertheless it still remains significantly quicker. However if it is instead given a series of axial scans ($\Pi_1$), then our method is just as quick as with $\Pi_2$, but also on average 41\% more accurate. RapidVol therefore prefers if US scans are acquired through longitudinal sweeps as opposed to rotational sweeps of the probe, both of which are readily available. Regardless of the acquisition technique, our method performs better when given more input scans. Since 2D US acquires images in real-time, collecting hundreds of images is quick and effortless, so the quality of the 3D reconstruction can easily be improved. Finally, our method generates novel views equally well in all three orthogonal planes, regardless of the acquisition technique used. Their accuracy is also much less variable from fetus to fetus than ImplicitVol (c.f standard deviations in \cref{tab:Accuracy comparison}).
    
\begin{figure}[b!]
\centering
\includegraphics[width=1\textwidth]{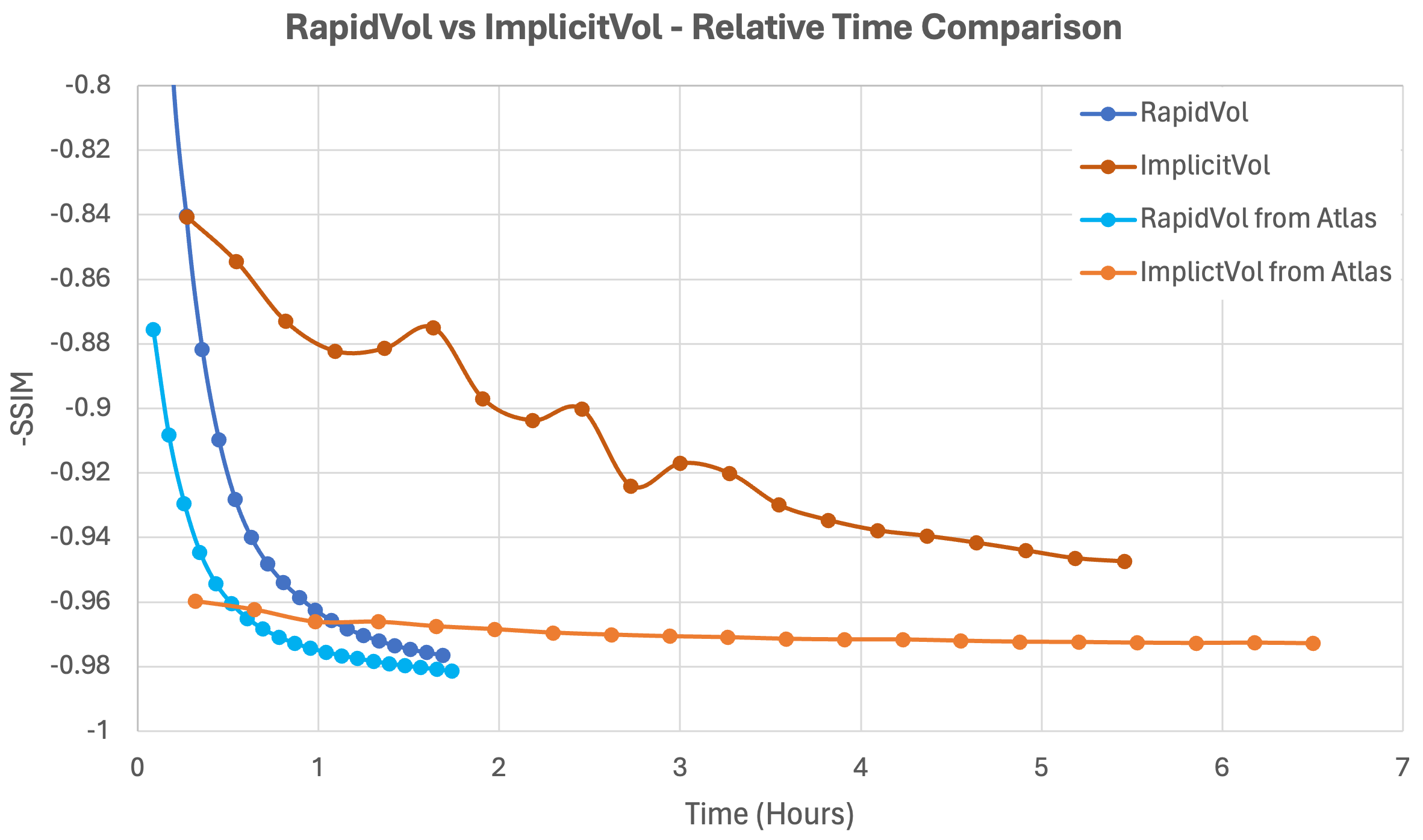}
\caption{Testing accuracy profile curves. Training was done to 5,000 epochs, accuracy reported every 250 epochs. For consistent timings, these runs were done on an isolated but \underline{much slower} GPU. $\Pi$ = 160 axial slices, Testing dataset = 160 coronal slices rotated $360^\circ$ about the Vertical Axis.}
\label{fig:RapidVol vs ImplicitVol - Time Trial}
\end{figure}

All runs were conducted on a shared high-performance cluster, however pre-emption by other processes often occured. Therefore to ensure a fair timing comparison between RapidVol and ImplicitVol, a selection of these runs were made on a much slower, but \textit{isolated} GPU (GTX TITAN X). The results of the \textit{relative} speed up, as well as the training profiles, can be seen in \cref{fig:RapidVol vs ImplicitVol - Time Trial}, showing that RapidVol is 3 times quicker than ImplicitVol per epoch. This significant speed-up, coupled with the accuracy improvements discussed above, makes for a very powerful reconstruction model. This experiment was run on a slow GPU, so upon deployment onto an isolated modern GPU, the absolute reconstruction/training time will be in the order of minutes as opposed to over 1 hour as shown in \cref{fig:RapidVol vs ImplicitVol - Time Trial}.\vspace{15pt}

\par{\noindent \textbf{Training from an Atlas:}}
\cref{fig:RapidVol vs ImplicitVol - Time Trial} shows that initialising from a pre-trained fetal brain atlas as opposed to randomly has no effect on reconstruction speed. It does however result in a far quicker rate of convergence, which is especially useful if the reconstruction process is required to be terminated after a certain duration (e.g. a 30 minute prenatal appointment). To create a reconstructed brain with a respectable SSIM of 0.9, initialising RapidVol from an atlas allows you to achieve this 2.7 times quicker than if RapidVol was initialised from random values. We therefore recommend all fetal brains are reconstructed using RapidVol which has been pre-initialised from a digital fetal atlas of the same age (these digital atlases are publicly available for most gestational ages \cite{namburete_normative_2023}).

\par{\noindent \textbf{Use of Inaccurate Poses:}}
\cref{tab:G.t Pose vs inaccurate Pose} shows that as expected, inaccurate poses lead to poorer reconstruction accuracy, regardless of the method. However when faced with inaccurate poses, which in practice will always be the case to some extent, our method is up to 35\% more accurate than the baseline. Even when a series of parallel sagittal poses are requested, we still perform 28\% better, demonstrating the robustness of our proposed method.

\begin{table}[h]
\caption{-SSIM accuracy results at test time, when our method does or does not have accurate initial poses. $\Pi$ = 256 coronal slices rotated $360^\circ$ about the Vertical Axis. Testing dataset = 256 \{axial, coronal, sagittal\} slices. More negative SSIM is better.}
\label{tab:G.t Pose vs inaccurate Pose}
\centering\scriptsize\setlength{\tabcolsep}{12pt}  %
\begin{tabular}{ccccc}
\toprule 
Poses & Method & Axial & Coronal & Sagittal\\
\midrule 
\multirow{2}{*}{Ground Truth} & RapidVol & -0.952 & -0.950 & -0.950\\
\cmidrule{2-5} 
 & ImplicitVol & -0.912 & -0.904 & -0.908\\
\midrule 
\multirow{2}{*}{Estimated} & RapidVol & -0.812 & -0.790 & -0.790\\
\cmidrule{2-5} 
 & ImplicitVol & -0.599 & -0.596 & -0.616\\
\bottomrule
\end{tabular}
\end{table}

\section{Conclusion}
We propose \textbf{RapidVol}, a hybrid implicit-explicit representation method utilising Tri-Planar Decomposition to rapidly reconstruct 3D ultrasound volumes from sensorless 2D scans. When compared to current state of the art, our method is up to 3x quicker, up to 46\% more accurate, and even if faced with slightly inaccurate poses is still robust, performing on average 32\% better that other methods do in that scenario. Reconstructing from a fetal atlas can also offer further speed-up. RapidVol helps to bring us one step closer to more efficient and accurate medical diagnostics, especially in settings where only basic 2D US probes are available but 3D views would be clinically beneficial.

\paragraph{Acknowledgements.} The authors acknowledge the generous support of the EPSRC Centre for Doctoral Training in Autonomous Intelligent Machines \& Systems (EP/S024050/1), Amazon Web Services, EPSRC Impact Acceleration Account Award, EPSRC Doctoral Prize Scheme, Royal Academy of Engineering (RF\textbackslash 201819\textbackslash 18\textbackslash 163), and the Bill \& Melinda Gates Foundation.

\newpage
\bibliographystyle{splncs04}
\bibliography{references-noURLs.bib}

\newpage
\setcounter{table}{0}
\renewcommand{\thetable}{A\arabic{table}}
\setcounter{equation}{0}
\renewcommand{\theequation}{A\arabic{equation}}
\section*{Appendix A - Positional Encoding} \label{sec:Appendix A}
Each element in feature vector $\mathbf{p}$ is encoded from $\mathbb{R}$ to $\mathbb{R}^{2L+1}$ by the following function:
\begin{equation}
    \gamma(p)=(p, \, \text{sin}(2^0 \pi p), \, \text{cos}(2^0 \pi p), \, \ldots ,\, \text{sin}(2^{L-1} \pi p), \, \text{cos}(2^{L-1} \pi p))
    \label{eq:Positional encoding}
\end{equation}
The degree of encoding (ie value of $L$), and whether or not to include p itself in $\gamma(p)$, are application specific choices.

\setcounter{table}{0}
\renewcommand{\thetable}{B\arabic{table}}
\setcounter{equation}{0}
\renewcommand{\theequation}{B\arabic{equation}}
\section*{Appendix B - Ablation Study} \label{sec:Appendix B}
\begin{table}[h]

\caption{Comparison between possible RapidVol setups. $L$ = positional encoding degree. ``+ input" = concatenate raw input with enocded input. Tri-Planar and CP Decomposition are both with Rank $R=5$, Channels $C=10$. Reconstruction is from 160 linearly spaced axial images. Testing dataset is 160 linearly spaced coronal slices. Values shown are reconstruction accuracy as measured with Negative SSIM (more negative is better). The best result for each network is shown in bold, with the overall best performance across the networks underlined. Network names are of the form ``MLP \textit{n}-\textit{w}". This stands for a Multilayer Perceptron (MLP) network with \textit{n} fully connected layers, and hidden layer(s) all of width \textit{w}. ReLU activation functions are used throughout, except for the last layer which has a sigmoid activation function. \vspace{5pt}}
\label{tab: Ablation results}

\centering\scriptsize\setlength{\tabcolsep}{6pt}  %
\begin{tabular}{cc|ccccccc}
\multicolumn{2}{c|}{\textbf{Network}} & $L=0$   & $L=2$   & $L=5$ & $L=10$ & \shortstack{$L=2$\\ + input} & \shortstack{$L=5$\\ + input} & \shortstack{$L=10$\\ + input} \\ 
\midrule
\multirow{7}{*}{\rotatebox[origin=l]{90}{Tri-Planar}}          & MLP 2-128           & -0.9568     & -0.9717      & -0.2573      & -0.2630     & \textbf{-0.9733}       & -0.2608             & -0.9303                      \\
                     & MLP 3-128           & -0.9647    & -0.9743      & -0.2591      & -0.3055       & \textbf{-0.9748}       & -0.2623             & -0.9243                    \\
                     & MLP 4-128           & -0.9709         & -0.9745      & -0.2617      & -0.8803      & \textbf{-0.9750}       & -0.2635             & -0.9403                \\
                     & MLP 3-32           & -0.9641       & -0.9744      & -0.2567      & -0.9341      & \textbf{-0.9758}       & -0.2636             & -0.9227                  \\
                     & MLP 3-64           & -0.9654     & -0.9738      & -0.2583      & -0.8738         & \textbf{-0.9743}       & -0.2632             & -0.9307                 \\
                     & MLP 2-64           & -0.9562     & -0.9729      & -0.2556      & -0.8448      & \textbf{\underline{-0.9764}}       & -0.2609             & -0.9167                    \\
\midrule
\multicolumn{2}{l|}{CP + MLP 2-64} & \textbf{-0.8544} & -0.8475      & -0.8265      & NaN    & -0.8512                & -0.8392             & -0.508                     
\end{tabular}
\end{table}

\end{document}